\documentclass[reprint,superscriptaddress,amsmath,amssymb,aps,prl,twocolumn]{revtex4-2}
\usepackage{graphicx}
\usepackage{dcolumn}\usepackage{bm}
\usepackage{hyperref}\hypersetup{colorlinks, citecolor=blue, linkcolor=blue, urlcolor=blue}
\usepackage[dvipsnames]{xcolor}\usepackage{geometry}
\begin{document}
	\title{Enhanced Cooper pairing in nano-patterned metals
	}
	\author{Masoud Mohammadi-Arzanagh}
	\email{masoudma@umd.edu}
	\affiliation{Joint Quantum Institute, Department of Physics,
		University of Maryland, College Park, MD 20742, USA}
	
	\author{Andrey Grankin}
	\author{Victor Galitski}
	\author{Mohammad Hafezi}

	\affiliation{Joint Quantum Institute, Department of Physics,
		University of Maryland, College Park, MD 20742, USA}
	
	\begin{abstract}
		Nano-patterning has been shown to be a powerful tool for manipulating the vibrational modes of elastic structures, with applications such as optical-mechanical mode coupling. Inspired by these recent developments in phononic band engineering, we propose a nano-patterning scheme to enhance the superconducting transition temperature $T_c$ in phonon-mediated nano-film superconductors, such as aluminum. Using the finite element method, we simulate the lattice vibrational modes of nano-patterned films within the Debye model. Our results show that periodic nano-patterning softens the lattice vibrational modes compared to bulk films. It also increases the density of states at high energies, resulting in a couple of percent enhancement in $T_c$. Moreover, we investigate connections to Weyl's law and provide an experimental design prescription to optimize nano-patterning for further enhancement of the superconducting transition temperature.
		
	\end{abstract}
	\maketitle
	\textit{Introduction---} Over the past couple of decades,  there has been tremendous progress in modifying phononic band structures using nano-fabrication and nano-patterning techniques~\cite{eichenfield2009optomechanical,safavi2010design,kippenberg2008cavity,eichenfield2009modeling,favero2009optomechanics,olsson2008microfabricated,vasseur2007waveguiding,gu2005design,mohammadi2008evidence}. Periodically applied defects of certain shapes induce modifications in lattice vibrations, which in turn lead to the formation of phononic crystals. Similar to the manipulation of light in photonic crystals, this advancement opens new possibilities for precisely controlling phonon modes as desired~\cite{eichenfield2009optomechanical}.
	A hallmark example of nano-patterning applications is the emergence of optomechanical crystals, where optical and mechanical mode coupling is enhanced using nano-scale periodic structures. Another example is the use of this technique to produce phononic band gaps in solids, which can suppress thermal vibrations and mitigate noise and dephasing in various physical processes \cite{florez2022engineering,sigalas1993band,kushwaha1993acoustic}. It can also be used to control and engineer the heat capacity and thermal conductance of a material. Depending on the patterning design, thermal conductance can be either suppressed or enhanced \cite{gorishnyy2005hypersonic,anufriev2022phonon,nomura2016phononic}.
	The ability to manipulate lattice vibrations through two-dimensional periodic patterning, enabling control over physical properties such as optomechanical coupling, heat capacity, thermal conductivity, and thermal noise, highlights the broad applicability of nano-patterning techniques. \\
	
	On another front, numerous experimental studies have demonstrated that granular superconductors, as well as thin films, often exhibit an enhanced superconducting transition temperature $T_c$~\cite{abeles1966enhancement,cohen1968superconductivity,strongin1968enhanced,prischepa2023phonon,ummarino2024quantitative}. This enhancement is attributed to the inherently disordered nature of granular superconductors. Specifically, the chaotic geometry of these grains has been shown to strengthen the effective electron-phonon coupling, thereby leading to a higher $T_c$~\cite{grankin2024enhanced}. This shows how nano-scale geometric features can enhance the transition temperature of elemental superconductors.\\ 
	\begin{figure}[b]
		\centering
		\includegraphics[scale=0.55]{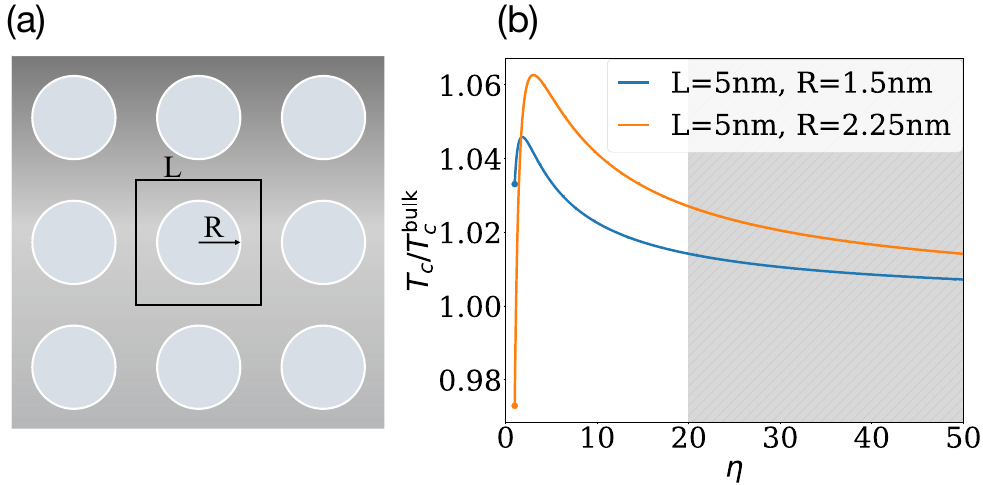}
		\caption{\textbf{(a)} The schematic shows an aluminum nanofilm with circular holes of radius $R$ being periodically patterned inside unit cells of side length $L$. \textbf{(b)} The enhancement in transition temperature as a function of the scaling factor $\eta$ for the two base geometries  $(L,R) = (5\,\text{nm},1.5\,\text{nm})$ and $(L,R) = (5\,\text{nm},2.25\,\text{nm})$. The perimeter-squared-to-area ratios for them are $\mathcal{L}^2/A = 4.95, 21.97$, respectively, where $\mathcal{L}=2\pi R, A=L^2-\pi R^2$. The base geometries are scaled such that $(L,R)\rightarrow (\eta  L,\eta R)$. 
			We observe an optimal scaling factor $\eta$ for each hole radius $R$, for which the sample exhibits the highest enhancement in transition temperature. This maximum enhancement in transition temperature is higher for the geometry with greater perimeter-squared-to-area ratio. 
			The gray dashed area represents the region where the size of the corresponding unit cell would be comparable to the superconducting coherence length~$\sim 100\text{nm}$, hence, the unit cell averaging scheme to obtain Eq.~\eqref{eq: el-funcation} would break down. The two dotted points at the beginning of each curve represent the $T_c$ values of unscaled geometries, i.e., $\eta = 1$, with transition temperatures of $T_c/T_c^{\text{bulk}} = 1.033$ for the blue curve ($R = 1.5\,\text{nm}$) and $T_c/T_c^{\text{bulk}} = 0.972$ for the orange curve ($R = 2.25\,\text{nm}$).}
		\label{fig:Fig1--}
	\end{figure}
	
	Motivated by such progress in nano-patterning and recent developments in the understanding of granular superconductors, it is intriguing to question whether modification of phononic band structures via nano-patterning can have an appreciable effect in phonon-induced superconductivity. Since nano-patterning length scales are usually lower bounded by 10nm, such schemes can only affect the low-energy acoustic phonons and naively one would expect that such patterning can not lead to any measurable effects. In this Letter, we investigate this question by examining the effect of nano-patterning on the lattice vibrational modes of elemental superconductors. To be concrete, we consider aluminum phononic band structure of a bulk nano-film and study its modification in the presence of nano-patterning within the Debye model. We study changes in the phononic density of states and demonstrate that nano-patterning softens the low energy phonon modes. Moreover, despite considering large-length-scale nano-patterning, we demonstrate that it can influence the entire phonon spectrum and observe an increased density of states at high energies for the patterned cases compared to the bulk, as predicted by the well-known Weyl-Vasilev law \cite{weyl1911asymptotische,vasil1986asymptotic}. We benchmark our findings for the density of states per unit area against the predictions of this law and further explore its connection to chaotic grains.  We then use these results to evaluate the Eliashberg function for both bulk and patterned samples and demonstrate that, for appropriate length scales, the high energy enhancement of this function in the patterned case can be used to obtain an enhancement of $\sim 6\%$ in $T_c$. It is worth emphasizing that the appropriate length scale can be determined by plotting the transition temperature as a function of the scaling factor $\eta$, i.e., $(L,R)\rightarrow(\eta L, \eta R)$, as shown in Figure~\ref{fig:Fig1--}. This allows us to identify the value of $\eta$ for which $T_c$ is maximized. Lastly, we propose a recipe to leverage the implications of Weyl-Vasilev law for optimizing fabrication patterns to further enhance the superconducting transition temperature.\\
	
	\textit{Geometry and Phonon Model---}To this end, we consider two scenarios: (1) a bulk two-dimensional aluminum nano-film, and (2) the same sheet of aluminum, nano-patterned with periodically arranged holes of circular shapes, as illustrated in panel (a) of Fig.~\ref{fig:Fig1--}. To analyze the phonons, we use the Debye model implying a linear spectrum with cut-off. We adjust this cut-off when we pattern and call it $\nu_D$ henceforth. Moreover, we ensure the hole features are much greater than the atomic scales  ~\cite{landau2012theory,hehl2002cauchy,marsden1994j}. The linear equation reads as:
	\begin{align}
		\rho \partial^2_t u_i = \sum_{jkl} C_{ijkl} \partial_j \partial_k u_l.
		\label{eq:phonon_wave}
	\end{align}
	Here, $\rho$ is the density of aluminum, $\vec{u}$ stands for the local lattice displacement, and $C_{ijkl}$ represents aluminum's stiffness tensor. Essentially, it is a fourth-rank tensor that connects the stress and strain within a linear elastic medium. In the isotropic  case, this tensor can be represented as:
	\begin{align}
		C_{ijkl}=\lambda_{\rm L} \delta_{ij}\delta_{kl}+\mu_{\rm L}(\delta_{ik}\delta_{jl}+\delta_{il}\delta_{jk})
	\end{align}
	where $\lambda_{\rm L}$ and $\mu_{\rm L}$ are the first and second Lamé parameters.
	For the second case, where holes are patterned in a periodic order, proper boundary condition at the hole boundary must be implemented to determine the phonon modes of Eq.~\eqref{eq:phonon_wave}. This condition, which is called free surface boundary condition, represents that the normal forces, both shear and compression, vanish at the boundary of aluminum and vacuum:
	\begin{align}
		\sum_{jkl} dS_j C_{ijkl} \partial_k u_l=0 \quad 	\forall i  \quad \in \quad {1,2}
		\label{eq: BC}
	\end{align}
	where $d\vec{S}$ is the normal differential length element around the hole boundary. Accompanied by Eq.~\eqref{eq:phonon_wave}, this boundary condition is sufficient to determine the spectrum of vibration modes. We used COMSOL Multiphysics to perform our finite element method simulations, solving Eq.~\eqref{eq:phonon_wave} and its associated boundary conditions.
	
	As well known for eigenvalues of Laplace operator, the boundary condition affects the asymptotic behavior of the density of states. In particular, Weyl's law establishes positive/negative correction for Neumann/Dirichlet boundary conditions, respectively  ~\cite{weyl1911asymptotische}. Weyl's law has been explored for Eq.~\eqref{eq:phonon_wave} with the free surface boundary condition, and the phonon spectrum exhibits a positive correction to the phonon density of states per unit area in a chaotic grain relative to the bulk case~\cite{weyl1911asymptotische,bertelsen2000distribution,vasil1986asymptotic}:
	\begin{align}
		\mathcal{N}(\nu) - \mathcal{N}_{\mathrm{bulk}}(\nu) = \frac{\beta_p \mathcal{L}}{2A c_s} \nu,
		\label{eq:weyl}
	\end{align}
	where $\mathcal{N}(\nu)$ is the number of states per unit area with energy below energy $h\nu$ for the grain, and $\mathcal{N}_{\text{bulk}}(\nu)$ is the corresponding quantity for the bulk geometry. Here, $\beta_p$ is an analytical constant with a value of $\beta_p = 2.085$ for aluminum~\cite{bertelsen2000distribution}. $\mathcal{L}$ represents the perimeter of the grain, and $A$ denotes its area. $c_s$ is the transverse sound velocity of aluminum, and $\nu$ is the cyclic frequency at which $\mathcal{N}$ is evaluated. In this work, we numerically demonstrate that the same expression holds for our periodically patterned hole geometry. Specifically, in our geometry Eq.~\eqref{eq:weyl} holds with $\mathcal{L}$ being the perimeter of the circle patterned inside the unit cell, and $A$ denotes the area of the unit cell excluding the area of the circle. Applying the Migdal-Eliashberg theory, we further show that this positive correction to the density of states can enhance the Eliashberg function and effective electron-phonon coupling. This, in turn, leads to an enhancement of the superconducting transition temperature.\\

	\textit{Electron-Phonon Coupling---}Having discussed the effect of nano-patterning on the phonon spectrum, we move on to addressing how this modification can be incorporated into the electronic aspect of the system. To achieve this, we need to introduce a model that captures the coupling of electrons to the lattice vibration modes. The simplest and most practical model for this purpose is the Hamiltonian introduced by Fröhlich~\cite{frohlich1954electrons,abrikosov2012methods,Altland:2010ww}:
	\begin{align}
		H=\sum_{l\vec{q}} E_{l,\vec{q}} 
		a^{\dagger}_{l,\vec{q}} a_{l,\vec{q}} +\sum_{k,\sigma=\uparrow,\downarrow} \xi_{\vec k} \psi^{\dagger}_{\sigma \vec k}\psi_{\sigma \vec k} \nonumber \\ +g\sum_{\sigma=\uparrow,\downarrow}\int \chi(\vec{r}) \psi^{\dagger}_{\sigma}(\vec{r})\psi_{\sigma}(\vec{r}) d\vec{r},
		\label{eq:Htotal}
	\end{align}
	where the first two terms represent the free Hamiltonians of the phonons and electrons, respectively, while the last term describes the coupling between them. $g$ represents the electron-phonon coupling strength, and $\chi(\vec{r})$ is defined as~\cite{abrikosov2012methods}:
	\begin{align}
		\chi(\vec{r})=\sum_{l\vec{q}}\frac{\nabla\cdot\vec{V}_{l,\vec{q}}}{\sqrt{2E_{l,\vec{q}}}}(a_{l,\vec{q}}+a^{\dagger}_{l,-\vec{q}}),
	\end{align}
	where $E_{l,\vec{q}}$ and $a^{\dagger}_{l,\vec{q}}(a_{l,\vec{q}})$ are the phonon energy and creation (annihilation) operator of the $l^{th}$ band with Bloch wave vector $\vec{q}$, i.e. $q_x,q_y \in  [-\pi/L,\pi/L]$ . $\xi_{\vec k}$ is the electron's dispersion and $\psi^{\dagger}_{\sigma}(\vec{r})$ is the creation operator of an electron at position $\vec{r}$. Mediated by phonons, the electrons experience an attractive potential, which pairs them up into the superconducting state below the superconducting transition temperature~\cite{cooper1956bound}. To find the critical temperature, we apply Migdal-Eliashberg theory, and estimate its solution for the transition temperature using McMillan's formula, where we perform the Fermi surface averaging of the phononic propagator
	\footnote{We use cyclic frequency throughout the work}:
	\begin{align}
		k_BT_c=\frac{h \Theta}{1.2}\exp\left[{-\frac{1.04(1+\lambda)}{\lambda-\mu^*(1+0.62 \lambda)}}\right].
		\label{eq:mcmillan}
	\end{align}
	Here, $k_B$ and $h$ are the Boltzmann and Planck constants, respectively. $\mu^*$ is a parameter representing the Coulomb repulsion, for which $\mu^*=0.1$ is a good approximation~\cite{allen1975transition, mcmillan1968transition}. It is worth noting that we do not take the effect of nano-patterning on $\mu^*$ into account. $\lambda$ and $\Theta$ are defined through electron phonon spectral function as the following:
	\begin{align}
		\lambda(\nu)=2\int_{0}^{\nu}\frac{\alpha^2F(\nu')}{\nu'}d\nu',
		\label{eq:lambda}
	\end{align}
	\begin{align}
		\zeta(\nu)=\frac{2}{\lambda(\nu_D)}\int_{0}^{\nu}\alpha^2F(\nu')d\nu',
	\end{align}
	where $\lambda=\lambda(\nu_D), \Theta=\zeta(\nu_D)$, and $\nu_D$ is the Debye cut-off frequency of the system in consideration. The electron phonon spectral function (Eliashberg function), $\alpha^2F(\nu)$, is the weight through which $\Theta$ and $\lambda$ are defined. It plays a key role in the determination of the transition temperature of superconductors, and provides a detailed description of the interaction between electrons and phonons, which is essential for accurately modeling superconducting properties~\cite{bose2009electron}. For the particular periodic geometry considered in this work this function reads as:
	\begin{align}
		\alpha^2 F(\nu)=\tilde{g}\sum_{ml} \int d^2 q \frac{1}{2E_{l,\vec{q}}} \frac{|\alpha_{l,\vec{q}}(\vec{K}_m)|^2\delta(h\nu-E_{l,\vec{q}})}{|\vec{q}+\vec{K}_m|\sqrt{1-\frac{|\vec{q}+\vec{K}_m|^2}{4k_F^2}}},
		\label{eq: el-funcation}
	\end{align}

	where we have defined:
	\begin{align}
		\tilde{g}=\frac{g^2 N(0)}{4\pi^2k_F}.    
	\end{align}
	Here, $E_{l,\vec{q}}$ is the phonon energy of the $l^{\text{th}}$ band at Bloch wave vector $\vec{q}$, and $\vec{K}_m = \frac{2\pi}{L}(m_x \hat{x} + m_y \hat{y})$ with $m_x, m_y$ being arbitrary integers. $k_F$ denotes the electronic Fermi wave number and $N(0)$ is the electronic density of states at the Fermi surface. $\alpha_{l,\vec{q}}(\vec{K}_m)$ are the Fourier coefficients of the periodic Bloch amplitude of the $\nabla \cdot \vec{V}_{l,\vec{q}}$ function, where $\vec{V}_{l,\vec{q}}(\vec{r})$ is the full phonon wave function of the $l^{\text{th}}$ band at Bloch wave vector $\vec{q}$. A detailed derivation of spectral function for our patterned geometry is provided in appendix~\ref{ap:A}.
	We determine $\tilde{g}$ by applying Eq.~\eqref{eq: el-funcation} to the bulk geometry, extracting the parameters required for McMillan's formula, and fitting $T_c$ to match the known transition temperature of bulk aluminum. Finding $\tilde{g}$ enables us to calculate the corresponding parameters for the patterned geometry, evaluate its transition temperature, and compare it with the bulk value.\\

    \begin{figure*}[t]
		\includegraphics[scale=0.9]{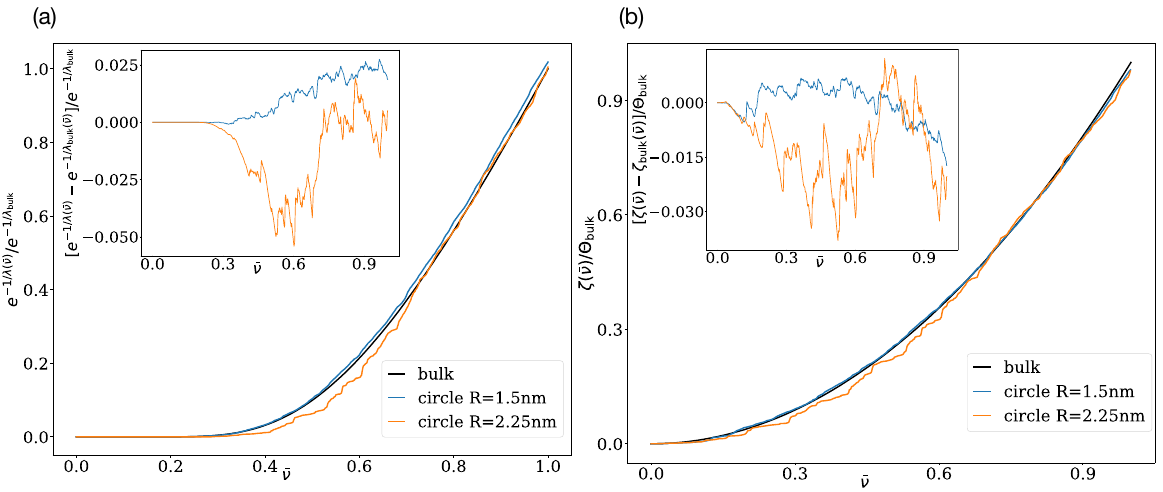}
		\caption{\textbf{(a)} The $e^{-1/\lambda(\bar{\nu})}/e^{-1/\lambda_\mathrm{bulk}}$ function for the unpatterned sample and the patterned samples with two different radii, $R = 1.5\,\text{nm}$ and $R = 2.25\,\text{nm}$, are plotted in black, blue, and orange, respectively, as a function of the dimensionless frequency ratio $\bar{\nu}=\nu/\nu_D$. The inset represents the difference of the blue and orange curve from the black in the entire domain of the plot. As it is more clear from the difference plot, the blue curve ends up having a larger Eliashberg parameter $\lambda$ than that of the bulk case $\lambda_{\mathrm{bulk}}$, while the orange does not show a sensible increase in this parameter compared to the bulk. \textbf{(b)} The corresponding $\zeta(\bar{\nu})/\Theta_{\mathrm{bulk}}$ function for each of the geometries discussed in panel (a) are plotted as a function of dimensionless frequency ratio $\bar{\nu}=\nu/\nu_D$. Once again, the inset represents the difference of the blue and orange curve from the black. Both of the curves have a smaller average frequency $\Theta=\zeta(\nu_D)$ than that of the bulk.}
		\label{fig:Fig2--}
	\end{figure*}

    \begin{figure}[h]
		\includegraphics[width=\linewidth]{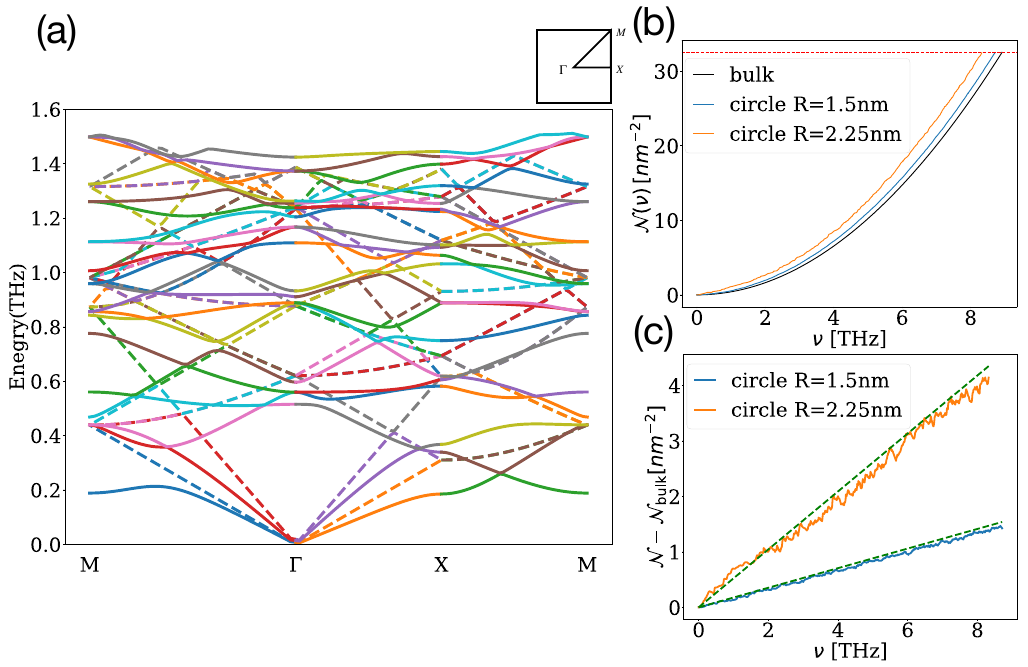}
		\caption{\textbf{(a)} The phononic band structure for both the bulk and nano-patterned samples, displaying the first 20 bands. Dashed lines represent the bulk sample, while solid lines correspond to the sample patterned with circular holes of radius $R = 1.5\,\text{nm}$ and a unit cell size of $L = 5\,\text{nm}$. Nano-patterning is observed to soften the phonon mode, thereby increasing the phonon density of states. The inset on the upper right corner illustrates the Brillouin zone with respect to which the band structure is plotted. \textbf{(b)} The phonon's cumulative density of states per unit area (i.e., the number of eigenstates below a certain energy) for the bulk sample and two patterned samples with radii $R = 1.5\,\text{nm}$ and $R = 2.25\,\text{nm}$. The dashed red horizontal line represents the value that corresponds to the total number of states per unit area for the bulk case; the Debye energy of the two other geometries are modified to stay consistent with this value, i.e., $\mathcal{N}_{\mathrm{bulk}}(\nu_D^{\mathrm{bulk}}) = \mathcal{N}(\nu_D)$. \textbf{(c)} The difference in cumulative density of states per unit area between the unpatterned sample and the two patterned samples with the same radii as in panel b, as a function of frequency. The dashed lines represent the analytic prediction for high energies based on Weyl-Vasilev law~\cite{bertelsen2000distribution}. 
			The geometry with a higher perimeter-squared-to-area ratio exhibits a greater enhancement of the density of states per unit area.}
		\label{fig:Fig3--}
	\end{figure}
    \textit{Results and Conclusion---} We begin the discussion of the results by benchmarking the modification of the density of states per unit area against the implications of Weyl-Vasilev law for high energies~\cite{weyl1911asymptotische,bertelsen2000distribution}. In particular, as shown in panel (c) of Fig.~\ref{fig:Fig3--}, we have plotted $\mathcal{N}(\nu) - \mathcal{N}_{\text{bulk}}(\nu)$ as a function of frequency. The linearity of this plot and its slope show good agreement with the analytic prediction of Eq.~\eqref{eq:weyl} at high energies, and the geometry with a greater perimeter squared-to-area ratio exhibits a higher enhancement of the density of states per unit area. This increase for high energies translates to an enhancement in the Eliashberg function at the same energy region as well, since both are affected by the same geometric factors. To be more precise, the density of states and two-point correlation functions of the eigenstates for the patterned geometry deviate from their bulk counterparts. Since the patterned geometry considered in this work is chaotic, the correction to these is comprised of two terms: a smooth (monotonic) term and a highly oscillatory term. The highly oscillatory contribution is sensitive to the nature of classical dynamics for a system whose classical counterpart is chaotic~\cite{garcia2011bcs,gutzwiller2013chaos}. This term is responsible for the oscillations observed in the density of states and Eliashberg function in this work. On the other hand, using the principles of random matrix theory and quantum chaos, one can show that the high energy part of the monotonic term in the cumulative Eliashberg function experiences a linear correction in frequency relative to the bulk case---similar to that of the cumulative density of states. This results from the fact that the high energy phonons can be modeled with a superposition of random and uncorrelated plane waves, entering from different directions with the same wavevector magnitude~\cite{brack1997semiclassical,kuhl2005classical} (see appendix ~\ref{ap:B} for further details).

    As evident from the behavior of $\lambda(\nu)$ and the Eliashberg function in panels (a) and (b) of Fig.~\ref{fig:FigA2--}, there is a competition between low-energy suppression and high-energy enhancement. The high-energy enhancement arises from the linear increase in both the cumulative Eliashberg function and the density of states relative to the bulk case at high frequencies~\cite{garcia2011bcs,gutzwiller2013chaos,brack1997semiclassical}.  Specifically, shapes with a higher perimeter to square root of area ratio exhibit a greater linear slope in the high-energy limit of the Eliashberg function as well. On the other hand, a smaller effective area, $L^2 - \pi R^2$, leads to a more extended low-energy suppression across the energy spectrum. These two effects compete, ultimately determining whether the electron-phonon coupling is enhanced or suppressed.

For the two specific patterned geometries considered in Fig.~\ref{fig:Fig2--}, $(L, R) = (5\,\text{nm}, 1.5\,\text{nm})$ (blue) and $(L, R) = (5\,\text{nm}, 2.25\,\text{nm})$ (orange), whose enhancements in $T_c$ correspond to the blue and orange dots in panel (b) of Fig.~\ref{fig:Fig1--}, the following observations can be made: For the blue curve $(L, R) = (5\,\text{nm}, 1.5\,\text{nm})$, despite the low-energy suppression, the high-energy enhancement dominates, resulting in an increase in the transition temperature. However, for the orange curve $(L, R) = (5\,\text{nm}, 2.25\,\text{nm})$, low-energy suppression prevails, leading to a decrease in transition temperature. Nevertheless, in what follows we show that scaling the size of the unit cell and circle proportionately can suppress the low-energy decrease in these functions and leverage the high-energy increase in them to enhance the transition temperature. Scaling to larger geometries can reduce the energy interval over which the mentioned low-energy decrease suppresses the Eliashberg function. This results in an overall enhancement of $T_c$ for both geometries, with the geometry having a greater perimeter-squared-to-area ratio showing an advantage due to a faster increase of the Eliashberg function at high energies.

	We continue the discussion of the results with a brief explanation of our scaling scheme. The linearity of the phonon equations implies that 
	$\mathcal{N}(\eta L, \eta R, \nu) \sim \mathcal{N}(L, R, \eta \nu)$. Assuming $1/k_F$ is much smaller than the other length scales in the problem, it also follows that 
	$\alpha^2 F(\eta L, \eta R, \nu) \sim \alpha^2 F(L, R, \eta \nu)$. This approach enables us to determine the density of states per unit area and the Eliashberg function for scaled geometries without the need to repeat the full phononic band structure simulation (see appendix ~\ref{ap:B} for further details).  As an example, considering the two scaled geometries with $(L,R)=(15\text{nm},4.5\text{nm})$ and $(L,R)=(15\text{nm},6.75\text{nm})$, we obtain the modified parameters $\lambda/\lambda_{\mathrm{bulk}}=1.011$, $\Theta/\Theta_{\mathrm{bulk}}=0.987$, $\nu_D/\nu_D^{\mathrm{bulk}}=0.992$ for the former case and
	$\lambda/\lambda_{\mathrm{bulk}}=1.019$, $\Theta/\Theta_{\mathrm{bulk}}=0.972$, $\nu_D/\nu_D^{\mathrm{bulk}}=0.979$ for the latter. These ratios translate to an enhancement of $\delta T_c/T_c=4.2\%$ and $6.3\%$ in the transition temperature for the two scaled geometries, respectively. A plot of transition temperature enhancement as a function of scaling factor is included in panel b of  Fig.~\ref{fig:Fig1--}; i.e. the geometry is scaled with a factor $\eta$ such that $(L,R)\rightarrow (\eta \times L,\eta \times R)$. This plot shows that there exists an optimal scaling factor $\eta$ for any base geometry for which $T_c$ is maximized, and the geometry with a higher perimeter squared-to-area ratio has a greater maximum enhancement. Having said that, it is clear that circles are not the best choice of patterning in terms of getting a greater transition temperature, and shapes like fractals that have higher perimeter-squared-to-area ratio would give even higher enhancement in $T_c$.

    However, it is crucial to consider that the fineness of the features in the geometry must not fall below the atomic scale of the underlying material. Due to computational complexities we do not include more complex shapes in the current work. Nevertheless, the goal of this Letter is to illustrate the trend in $T_c$ enhancement and to demonstrate the significant role that nano-patterning plays in improving the superconducting transition temperature of 2D materials, such as aluminum nano-films.

	In this work, we focused on simple circular hole patterning. A clear direction for future work is to optimize the patterning shape and size to further enhance \( T_c \), using machine learning techniques~\cite{lopez2023self}. In addition, a more realistic model for phonons could be developed by accounting for the lattice structure of the underlying material and solving for the phonon modes. This approach would automatically handle the Debye cut-off frequency. Other exciting directions for future research include replacing aluminum with other phonon-mediated superconductors, such as niobium, and exploring the potential of inducing superconductivity in non-superconducting materials.
	
	\textit{Acknowledgment---} This work was supported by 
	DARPA HR001124-9-0310.

	\bibliography{Refrences}
	\appendix
	
	\section{The derivation of Eliashberg function}
	\label{ap:A}
	In the absence of translation invariance, the phonon propagator and the two body electron-electron interaction, which is mediated by phonons, are not necessarily functions of coordinated difference. Therefore, we can write the following general form for the phonon mediated electron-electron interaction:
	\begin{align}
	D(r,r';\epsilon_n)=g^2\sum_{l\vec{q}}\frac{1}{2 E_{l,\vec{q}}}\phi_{l,\vec{q}}(\vec{r})\phi^*_{l,\vec{q}}(\vec{r}')\frac{2 E_{l,\vec{q}}}{E_{l,\vec{q}}^2+\epsilon_n^2}
	\end{align}
	Here, $g$ is a dimensionful constant determining the electron-phonon coupling strength, $E_{l,\vec{q}}$ is the energy of the $l^{th}$ band with Bloch wave vector $\vec{q}$, and $\phi_{l,\vec{q}}(\vec{r})$ is the divergence of the corresponding phonon wave function:
	\begin{align}
		\phi_{l,\vec{q}}(\vec{r})=\nabla \cdot \vec{V}_{l,\vec{q}}(\vec{r}),
	\end{align}
	where $\vec{V}_{l,\vec{q}}$s are the full phonon wave functions for $l^{th}$ band and Bloch wave vector $\vec{q}$. Using Bloch's theorem we can write the function $\phi_{l,\vec{q}}(\vec{r})$ in the following form:
	\begin{align}
		\phi_{l,\vec{q}}(\vec{r})=\varphi_{l,\vec{q}}(\vec{r})e^{i \vec{q} \vec{r}}.
	\end{align}
	Once again, $\vec{q}$ is the Bloch wave vector in the first Brillouin zone, and $\varphi_{l,\vec{q}}(r)$ is the periodic amplitude that can be expanded in the Fourier basis:
	\begin{align}
		\varphi_{l,\vec{q}}(\vec{r})=\sum_m \alpha_{l,\vec{q}}(\vec{K}_m)e^{i\vec{K}_m.\vec{r}}.
	\end{align}
	Here, $\vec{K}_m=\frac{2\pi m_x}{L}\hat{x}+\frac{2\pi m_y}{L}\hat{y}$ with $m_x,m_y$ being arbitrary integers and $L$ being the side length of the unit cell.
	Substituting this in the phonon propagator and expressing it in terms of the relative $\vec{r}-\vec{r}'=\vec{\ell}$ and center of mass coordinates $\vec{R}=\frac{\vec{r}+\vec{r}'}{2}$ yields:
	\begin{align}
		&D(\vec{R},\vec{\ell};\epsilon_n)=g^2\sum_{l\vec{q}}\frac{1}{2 E_{l,\vec{q}}}\phi_{l,\vec{q}}(\vec{r})\phi^*_{l,\vec{q}}(\vec{r}')\frac{2 E_{l,\vec{q}}}{E_{l,\vec{q}}^2+\epsilon_n^2}\nonumber \\ 
		&=g^2\sum_{lkmm'} \frac{1}{E_{l,\vec{q}}^2+\epsilon_n^2}\alpha_{l,\vec{q}}(\vec{K}_m)\alpha^*_{l,\vec{q}}(\vec{K}_m')\nonumber \\
		&\quad e^{i[\vec{q}+\frac{1}{2}(\vec{K}_m+\vec{K}_{m'})]\cdot\vec{\ell}}e^{i(\vec{K}_{m'}-\vec{K}_m)\cdot\vec{R}}.
	\end{align}
	This expression does not have translation invariance, but, averaging over the unit cell removes the dependence on center of mass coordinate $\vec{R}$ and the result is only a function of difference. This is a justified step due to the fact that superconducting coherence length is much larger than the unit cell size that we are considering in this work. Performing the integration over $\vec{R}$ then yields:
	\begin{align}
		&\bar{D}(\vec{\ell},\epsilon_n)=g^2\sum_{l\vec{q}\vec{K}_m} \frac{1}{E_{l,\vec{q}}^2+\epsilon_n^2}\big|\alpha_{l,\vec{q}}(\vec{K}_m)\big|^2e^{i(\vec{\vec{q}}+\vec{K}_m).\vec{\ell}}\nonumber \\&=\sum_{\vec{k}} \bar{D}(\vec{k}, \epsilon_n)e^{i \vec{k}.\vec{\ell}},
	\end{align}
	where $\vec{k}= \vec{q}+\vec{K}_m$ 
	is the momentum that electrons gain via scattering off of phonons. 
	We use this result and combine it with the already existing expression for the Eliashberg function in the literature~\cite{bose2009electron}, to get the Eliashberg function for our geometry:
	\begin{align}
		\alpha^2 F(\nu)=\tilde{g}\sum_{ml} \int d^2 q \frac{1}{2E_{l,\vec{q}}} \frac{|\alpha_{l,\vec{q}}(\vec{K}_m)|^2\delta(h\nu-E_{l,\vec{q}})}{|\vec{q}+\vec{K}_m|\sqrt{1-\frac{|\vec{q}+\vec{K}_m|^2}{4k_F^2}}}, 
	\end{align}
	where:
	\begin{align}
		\tilde{g}=\frac{g^2 N(0)} {4\pi^2k_F}.   
	\end{align}
	Here, $k_F$ is the Fermi wave vector, and $N(0)$ is the electronic density of states at the Fermi surface. We evaluate the above expression by solving for the phonon spectrum using \textit{COMSOL Multiphysics} to obtain $V_{l,\vec{q}}(r)$, and then taking the Fourier transform of its divergence to find $\alpha_{l,\vec{q}}(\vec{K}_m)$.\\
 
\section{Numerical fitting for high energy part of $\alpha^2F(\nu), \lambda(\nu)$}
\label{ap:B}

Following arguments from random matrix theory and quantum chaos, we aim to justify the linear behavior of the Eliashberg function at high energies. To analyze the high-energy behavior of the monotonic part of the Eliashberg function, we start from the following expression:
\begin{align}
    &\alpha^2F(\nu)=\nonumber \\& \tilde g \sum_{l,\vec q}\int d^2k\frac{ \langle \phi_{l,\vec q}(\vec R+\vec \ell/2)\phi^*_{l,\vec q}(\vec R-\vec \ell/2)\rangle_{\vec R} (\vec k)\delta(h\nu-E_{l\vec q})}{2kE_{l\vec q}\sqrt{1-\frac{k^2}{4k_F^2}}},
\end{align}
where $\langle \cdot \rangle_{\vec R}$ represents a center-of-mass unit-cell average, and $\langle \cdot \rangle_{\vec R}(\vec k)$ is the corresponding Fourier transform with respect to $\vec \ell$ at wavevector $\vec k$.

We then use the \textit{random plane wave approximation} (also known as Berry’s conjecture in the quantum chaos literature) to express the divergence of the wavefunction ($\phi_{l,\vec q}$) at high energies as a superposition of plane waves with random and uncorrelated amplitudes~\cite{brack1997semiclassical,kuhl2005classical}. This leads to the following ensemble average:
\begin{align}
    \overline{\langle \phi_ {l,\vec q}(\vec R+\vec \ell/2)\phi^*_{l,\vec q}(\vec R-\vec \ell/2)\rangle_{\vec R}}(\vec k)
    \simeq \Gamma_{l,\vec q}|\vec k|\delta\left(|\vec k| - \frac{E_{l\vec q}}{\hbar c_{p}}\right),
\end{align}
where the overline indicates ensemble averaging, $c_p$ is the longitudinal sound velocity, and $\Gamma_{l,\vec q}$ is a parameter that determines the density of states for longitudinal phonons at high energies.

Substituting this into the expression for the Eliashberg function gives:
\begin{align}
    \alpha^2F(\nu)\propto\frac{1}{\sqrt{1-\frac{\pi^2\nu^2}{c_{p}^2k_F^2}}}\frac{d\mathcal{N}_{\parallel}}{d\nu},
\end{align}
where $\mathcal{N}_{\parallel}(\nu)$ is the cumulative density of states of longitudinal phonons, and $c_p$ is corresponding sound velocity. At high energies, $\mathcal{N}_{\parallel}(\nu)$ deviates linearly with frequency from the bulk behavior. Assuming $1/k_F$ is much smaller than other relevant length scales, we find that the correction to the cumulative Eliashberg function at high energies relative to the bulk is linear in frequency~\cite{brack1997semiclassical,garcia2011bcs,gutzwiller2013chaos,kuhl2005classical,grankin2024enhanced}:
\[
\int_0^{\nu}\big[\alpha^2F(\nu') - \alpha^2F_{\mathrm{bulk}}(\nu')\big] d\nu' = s \nu+ r,
\]
where $s$ is the slope of the linear correction, and $r$ is the y-intercept.

\begin{figure}[h]
    \includegraphics[width=\linewidth]{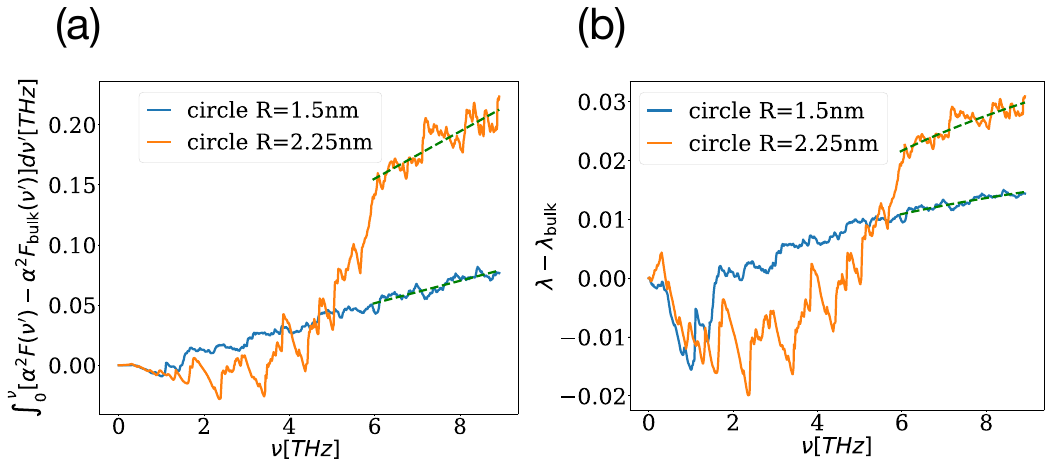}
    \caption{\textbf{(a)} The difference in the Eliashberg function between the bulk and patterned samples with radii of $R = 1.5\,\text{nm}$ and $R = 2.25\,\text{nm}$, as a function of frequency. The high-energy region of both curves exhibits linear behavior, with the slope determined via numerical linear fitting. This linear fit is utilized to estimate the Eliashberg function for similarly scaled geometries, i.e., geometries where all lengths are scaled while keeping their ratios fixed. \textbf{(b)} The difference in $\lambda$ between the bulk and patterned samples, with radii of $R = 1.5\,\text{nm}$ and $R = 2.25\,\text{nm}$, plotted as a function of frequency. Given the linear behavior of the difference in the Eliashberg function at high energies in panel a, the high-energy part of both curves in this plot is expected to exhibit logarithmic behavior. The dashed green curves represent the logarithmic numerical fit. Like in panel a, these fits are used to estimate $\lambda$ for scaled geometries where all lengths are increased proportionally while keeping their ratios fixed.}
    \label{fig:FigA2--}
\end{figure}

Consequently, the high-energy part of \(\lambda(\nu) - \lambda_{\text{bulk}}(\nu)\) is expected to exhibit logarithmic behavior. In panels (a) and (b) of Fig.~\ref{fig:FigA2--}, we plot these deviations from the bulk values along with their numerical fits. The slope of the linear fit and the parameters of the logarithmic fit were extracted using numerical fitting packages.

Alternatively, one could use the Green's function approach outlined in the literature~\cite{garcia2011bcs,gutzwiller2013chaos} to analytically derive the slope. However, to keep the math simple, we do not include it here.

In fact, the cumulative Eliashberg function \(\int_{0}^{\nu} \alpha^2F(\nu')\,d\nu'\) and \(\lambda(\nu)\) both exhibit a positive high-energy correction relative to the bulk in the patterned samples. As shown in Fig.~\ref{fig:FigA2--}, the geometry with a larger perimeter-squared-to-area ratio displays a greater increase in both functions at high energies.  The numerical fits (green dashed curves) allow us to extrapolate these functions and find the corresponding values at any higher energy of interest. These values are used to find the Eliashberg parameters of scaled geometries. By scaling to larger geometries, the enhancement of these functions at high energies can be leveraged to counteract the low energy suppression.

\section{Downward energy level shift induced by patterning}
\label{ap:C}

According to Weyl's law for our chaotic system, the density of states at high energies deviates from the bulk counterpart by a constant term. This results in a uniform downward shift of the high-energy spectrum on average, i.e., 
\[
E^{\text{patterned}}_{l,\vec q} \simeq E^{\text{bulk}}_{l,\vec q} - \Delta E.
\]
To demonstrate this, we begin with the following generic expression for the cumulative density of states:
\begin{align}
\mathcal{N}(\nu) &= \frac{1}{(2\pi)^2} \sum_l \int_{\text{BZ}} \Theta(h\nu - E^{\text{bulk}}_{l\vec q} + \Delta E_{l\vec q}) \, d^2q \nonumber \\
=& \mathcal{N}_{\text{bulk}}(\nu) + \frac{1}{(2\pi)^2} \sum_l \int_{\text{BZ}} \frac{d}{hd\nu} \Theta(h\nu - E^{\text{bulk}}_{l\vec q}) \Delta E_{l\vec q} \, d^2q,
\end{align}
where we have Taylor-expanded the Heaviside function to linear order in $\Delta E_{l\vec q}$ to extract the leading-order correction.

Assuming a constant energy shift at high energies $\Delta E_{l\vec q} = \Delta E$, the expression simplifies to:
\begin{align}
\mathcal{N}(\nu) = \mathcal{N}_{\text{bulk}}(\nu) + \frac{d}{hd\nu} \mathcal{N}_{\text{bulk}}(\nu)  \Delta E.
\end{align}

Using the known bulk expression $\mathcal{N}_{\text{bulk}}(\nu) = \frac{\pi\nu^2}{c_s^2}(1 + 1/\kappa^2)$, where $\kappa = c_p/c_s$ is the ratio of longitudinal to transverse sound velocity, we arrive at the Weyl's formula for the patterned geometry at high energies. This justifies the assumption of a constant energy shift. Equating the proportionality factors gives:
\begin{align}
\Delta E = \frac{h c_s \beta_p \mathcal{L}}{4\pi A(1 + 1/\kappa^2)}.
\end{align}

The energy spectrum of the patterned system is softened at low energies, meaning each level is shifted downward on average. This energy shift saturates to the above value as we move to higher-energy bands.
Roughly speaking, this can be viewed as a mapping from a given patterned state to a higher-energy bulk state. Given the monotonically increasing behavior of the cumulative density of states and cumulative Eliashberg function in the bulk, this mapping explains the enhanced values of these functions in the patterned system relative to the bulk.
Thus, even though the enhancement in the Eliashberg function and density of states due to Weyl's law becomes prominent at high energies, the effect originates from modifications to the spectrum at low energies.

That being said, adopting a more realistic lattice model, which modifies the Debye phonon spectrum, would not eliminate the enhancement of the Eliashberg parameters—hence, the observed increase in $T_c$ remains robust.
 
\section{Numerical implementation details}
\label{ap:D}
We performed the phonon simulations using \textit{COMSOL Multiphysics}, within the \textit{Solid Mechanics} module in a two-dimensional (2D) geometry. The \textit{Eigenfrequency} study type was used to compute the vibrational modes.
For both the patterned and unpatterned (bulk) cases, we constructed a square unit cell of aluminum. In the patterned case, a circular hole of the desired radius was created within the unit cell.  A physics-controlled mesh was applied, with the mesh resolution chosen based on the geometry and the relevant energy scales of the problem.
\begin{table}[h!]
\centering
\caption{Material parameters and values used in COMSOL simulations}
\begin{tabular}{l c}
\hline
Parameter & Value/Type \\
\hline
Elasticity module (E) & $\simeq 70$ GPa \\
Poisson ratio & $\simeq 0.33$ \\
Unit cell length (L) & 5 nm \\
Patterned hole radius & 2.25 nm, 1.5 nm \\
Aluminum density & $\simeq 2.7$ g/cm$^3$ \\
$T_c$ of bulk aluminum & $\simeq 1.2$ K \\

Lattice model & Isotropic \\
\hline
\end{tabular}
\label{tab: numvals}
\end{table}

To account for the periodicity of the system, we imposed \textit{Floquet periodic boundary condition} in \textit{COMSOL Multiphysics}  (Bloch theorem is applied) which allows specifying the Bloch wavevector in the Brillouin zone and run the simulation for that specific value. The simulations were repeated for a range of two-dimensional (2D) Bloch momentum values to compute the phonon spectrum throughout the entire Brillouin zone.

For the patterned case, the boundaries around the hole were set to \textit{free surface boundary condition}, corresponding to zero net force on the boundary points. \\
The numerical values used in our simulations and calculations are listed in TABLE ~\ref{tab: numvals}.
\section{Brillouin zone mesh convergence}

We have already discussed that the highly oscillatory behavior observed in the Eliashberg functions, $\lambda$, and the density of states originates from the chaotic nature of the system. To verify that these features are not numerical artifacts arising from insufficient meshing in our finite element method simulations, we perform convergence tests using two different Brillouin zone meshes. 

\begin{figure}[h] 
    \centering
    \includegraphics[width=\linewidth]{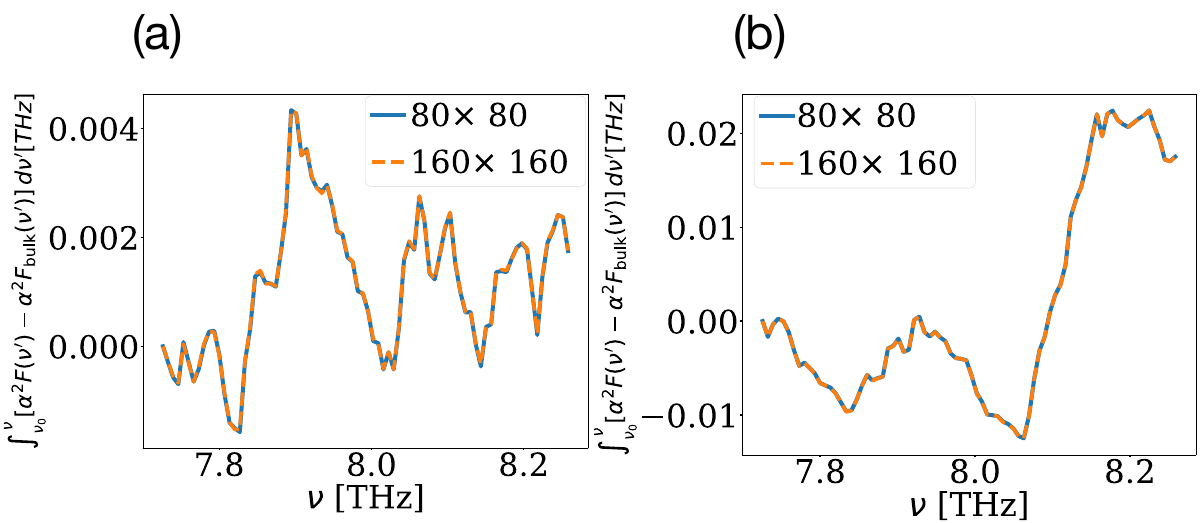}
\caption{
\textbf{(a)} Comparison of cumulative Eliashberg functions relative to the bulk for two Brillouin zone meshes, for the geometry $(L,R) = (5\,\text{nm}, 1.5\,\text{nm})$. The blue solid line represents results obtained using an $80 \times 80$ $q$-point mesh, while the orange dashed line corresponds to $160\times 160$ mesh. \textbf{(b)} Same comparison for the $(L,R) = (5\,\text{nm}, 2.25\,\text{nm})$ geometry. In both cases, the reference frequency is $\nu_0 \simeq 7.73\,\text{THz}$, and the results demonstrate convergence with respect to Brillouin zone sampling.
}

    \label{fig:mesh_test}
\end{figure}

Figure~\ref{fig:mesh_test} shows the difference between the cumulative Eliashberg functions for the two patterned cases relative to the bulk case:
\begin{align}
    \int_{\nu_0}^{\nu} \left[\alpha^2F(\nu') - \alpha^2F_{\text{bulk}}(\nu')\right] \, d\nu'.
\end{align}
Here, $\nu_0$ is an arbitrary reference frequency from which the cumulative integral is computed.
In these simulations, we impose Floquet-periodic boundary conditions on the unit cell, which ensures that the wavefunctions satisfy Bloch’s theorem. We then compute the phonon spectrum for different Bloch momenta across the Brillouin zone. By repeating the simulation at each $q$-point and collecting the resulting data, we obtain the Eliashberg function and the phonon density of states.
While increasing the mesh density in the Brillouin zone improves resolution, it also significantly increases the computational cost due to the need to sample the entire zone. To check mesh adequacy, we choose a high-frequency interval beginning, for example, at $\nu_0 \simeq 7.73\,\text{THz}$ and perform the phonon simulations accordingly. It is worth noting that the specific choice of $\nu_0$ does not carry any physical significance; any other high-frequency reference point could equally be used.

As shown in Fig.~\ref{fig:mesh_test}, increasing the number of $q$-points does not significantly change the integrated Eliashberg function, which confirms the validity of the original mesh used in our main simulations.

\end{document}